\documentclass[aps,prd,showpacs,preprintnumbers,twelvepoint,floatfix]{revtex4}
\usepackage{epsfig}
\usepackage{latexsym,amsmath,amssymb,amstext}
\usepackage{float}
\usepackage{graphicx}
\usepackage{color}

\newcommand\bef{\begin{figure}}
\newcommand\eef[1]{\label{fg:#1}\end{figure}}
\newcommand\beq{\begin{equation}}
\newcommand\eeq[1]{\label{#1}\end{equation}}

\newcommand\fgn[1]{Figure \ref{fg:#1}}
\newcommand\eqn[1]{eq.\ (\ref{#1})}
\newcommand\scn[1]{Section \ref{sec:#1}}
\newcommand\apx[1]{Appendix \ref{sec:#1}}

\newcommand\ie{{\sl i.e.\/}}

\newcommand\etal{{\sl et al.\/}}
\newcommand\jhep{{\sl J.\ H.\ E.\ P.\/}\ }

\newcommand\sun{\rm{SU}\left(N_c\right)}
\newcommand\Sigmab{\overline{\Sigma}}
\newcommand\tr{{\rm Tr}}
\newcommand\re{{\rm Re}}
\newcommand\dbd[2]{\frac{d #1}{d #2}}
\newcommand\del[2]{\frac{\partial #1}{\partial #2}}
\newcommand\pa[1]{\left[#1\right]_{\rm TA}}
\newcommand\deriv[3]{\left[\frac{\delta #1}{\delta #3}\right]_{#2}}
\newcommand\delt[3]{\dbd{#1}{#3}\bigg|_{#2}}

\begin{document}
\title{A Concise Force Calculation for Hybrid Monte Carlo with Improved Actions}
\author{Nikhil\ \surname{Karthik}}
\email{nikhil@theory.tifr.res.in}
\affiliation{Department of Theoretical Physics, Tata Institute of Fundamental
         Research,\\ Homi Bhabha Road, Mumbai 400005, India.}
\begin{abstract}
We present a concise way to calculate force for Hybrid Monte Carlo with
improved actions using the fact that changes in thin and smeared link
matrices lie in their respective tangent vector spaces. Since hypercubic
smearing schemes are very memory intensive, we also present a memory
optimized implementation of them.

\end{abstract}
\preprint{TIFR/TH/14-01}
\maketitle
\section{Introduction}\label{sec:intro}
A  standard method for dynamical simulation of QCD is by simulating an
equivalent micro-canonical ensemble of a  fictitious classical system
\cite{polonyi}.  Another method for global updates of gauge fields is
through stochastic evolution using Langevin dynamics \cite{wilson}.
However, in both these methods, the finite step size used for the
evolution through the simulation time introduces error. Hybrid Monte Carlo
(HMC) \cite{hmc} avoids this problem by combining molecular dynamics
with Langevin-type refreshment of the momenta conjugate to the gauge
links at the beginning of every trajectory and an acceptance step in the
end. Such a micro-canonical evolution using classical dynamics requires
the calculation of force, which in this case is the derivative of the
Hamiltonian with respect to the gauge links.

Nowadays, gauge link smeared actions are commonly used in dynamical
simulations to improve scaling behaviour, especially after the
advent of stout \cite{ST} and HEX \cite{HEX} schemes due to their
differentiability. These actions are explicit functions of the smeared
links. A method to calculate force for HMC with these improved
actions was first discussed in \cite{kamleh}, where a chain rule was used
to find the variation of action with respect to the original thin-links.
Using this chain rule, the force for HEX improved fermions was derived
in \cite{bmw}, combining the methods utilised in \cite{hfratz} (for the
``hypercubic" part of HEX) and \cite{ST} (for the ``$\exp$" part of HEX).

In this work, we show the simplicity of finding the force using the
ordinary derivative with respect to a single real parameter, which is the
magnitude of displacement in the tangent vector space of a gauge link.
We take this method further to re-derive the force for the HEX improved
HMC. For this, we note that a variation in a thin-link, which lies in
its tangent vector space, causes variations in smeared links which also
lie in their respective tangent vector spaces.  With this observation,
we again have to find directional derivatives with respect to a single
real parameter in each of the tangent vector spaces of the smeared links.
The unknowns are the directions in these tangent vector spaces, which are
relatively  easy to find.  This method differs substantially from the
one used in \cite{bmw} by not requiring to find derivatives of action
with respect to each matrix element of thin and smeared links. We find
that this makes the calculation easy to check and that it can be easily
extended to other nested improvement schemes, which would otherwise be
difficult due to the constant book-keeping of colour indices required.

This work is based on the calculation of HEX force in \cite{bmw} and we
borrow their notation as much as possible. In \scn{Def}, we introduce
our notation and describe the construction of HEX smeared quarks.
\scn{tangent} deals with finding the derivatives of functions defined
on an $\sun$ manifold. Using the methods developed, we give a concise
derivation of the HEX force in \scn{force}.  In \scn{memopt}, we give
a memory optimized implementation of hypercubic smearing schemes.

\section{Definitions}\label{sec:Def}
The lattices contain $N_t$ points along the temporal direction and
$N_s$ points in the spatial directions. The volume of the lattice
is $V_{lat}=N_t N_s^3$.  A point on this lattice is labelled by
the variables $x, y$ and $l$.  The directions are labelled by Greek
indices and their values run from 1 to 4.  In this notation, the gauge
link connecting a point $x$ to $x+\mu$ is written as $U_{x,\mu}$. The
Einstein summation convention is not used and summations over indices
are indicated explicitly.  To be concrete, we deal with $N_f$ flavours of
staggered quarks.  The standard staggered Dirac operator is constructed
out of thin-links, $U$, and it is given by
\beq
D_{x,y}=m\delta_{x,y}+\frac{1}{2}\sum_{\nu=1}^4\left( U_{x,\nu}\delta_{x,y-\nu}-U^\dagger_{x-\nu,\nu}\delta_{x,y+\nu}\right),
\eeq{stagdef}
where the gauge links have been pre-multiplied with the staggered phases,
$\eta_{x,\nu}$, and $m$ is the bare quark mass in lattice units.

A single level of HEX has three nested sub-levels constructed such that
the smeared links at the $n$-th sub-level, $V^{(n)}$, are built only
out of the thin-links within a hypercube. The final HEX smeared links,
$V^{(3)}_\mu$ obtained in the third sub-level, are given by \cite{HEX}
\beq
\begin{split}
\text{\rm Sub-level 1:}{}& \quad V^{(1)}_{x,\mu;\nu\rho}=\exp\left(\pa{C^{(1)}_{x,\mu;\nu\rho}U^\dagger_{x,\mu}}\right)U_{x,\mu},\qquad C^{(1)}_{x,\mu;\nu\rho}=\frac{\epsilon_1}{2}\sum_{\pm\sigma\ne\mu\nu\rho}\left[U_{x,\sigma}U_{x+\sigma,\mu}U^\dagger_{x+\mu,\sigma}\right]\\
\text{\rm Sub-level 2:}{}& \quad V^{(2)}_{x,\mu;\nu}=\exp\left(\pa{C^{(2)}_{x,\mu;\nu}U^\dagger_{x,\mu}}\right)U_{x,\mu},\qquad  C^{(2)}_{x,\mu;\nu}=\frac{\epsilon_2}{4}\sum_{\pm\sigma\ne\mu\nu}\left[ V^{(1)}_{x,\sigma;\mu\nu} V^{(1)}_{x+\sigma,\mu;\sigma\nu}V^{(1)\dagger}_{x+\mu,\sigma;\mu\nu}\right]\\
\text{\rm Sub-level 3:}{}&\quad V^{(3)}_{x,\mu}=\exp\left(\pa{C^{(3)}_{x,\mu}U^\dagger_{x,\mu}}\right)U_{x,\mu},\qquad C^{(3)}_{x,\mu}=\frac{\epsilon_3}{6}\sum_{\pm\sigma\ne\mu}\left[ V^{(2)}_{x,\sigma;\mu} V^{(2)}_{x+\sigma,\mu;\sigma}V^{(2)\dagger}_{x+\mu,\sigma;\mu}\right],
\end{split}
\eeq{hxdef}
where $\pa{\ldots}$ is the traceless anti-hermitean part of its
argument. Also, none of the directional indices are equal to each
other.  $C^{(n)}$ are staples constructed out of the smeared links in
the $(n-1)$-th sub-level weighted by the tunable smearing parameters,
$\epsilon_n$. Dropping the directional indices for the sake of brevity,
as will be done quite often in \scn{force}, \eqn{hxdef} can be written
in short as
\beq
V^{(n)}_x=\exp(A^{(n)}_x)U_x \quad\text{\rm with}\quad A^{(n)}_x=\pa{C^{(n)}_xU^\dagger_x}.
\eeq{hexschematic}
The above construction is devoid of non-analytic operations, like
projection to $\sun$. This means that $V^{(3)}$ can be expanded as a power
series in $U$, however complicated the resulting expression might be.
The one level HEX improved Dirac operator is given by replacing $U$
with $V^{(3)}$ in \eqn{stagdef}.

\section{Manifold, tangent vector space and derivatives}\label{sec:tangent}
In this section, we state some of the required results in Lie groups
in the context of $\sun$. We refer the reader to \cite{daniel,gilmore}
for an extensive introduction to this topic.  The group $\sun$ forms
an $N=N_c^2-1$ dimensional manifold, with each point, $U$, on it being
a group element. This means that there is a mapping, $\phi$, from the
neighbourhood of $U$ to $\mathbb{R}^{N}$, called a coordinate chart.
Let $f(U)$ be a real valued function defined over the manifold.
We use the same notation $f$ to refer to both $f$ defined on the manifold
as well as $f\circ\phi^{-1}$ defined on the coordinate chart.

We are interested in the fundamental representation of $SU(N_c)$, in
which case, points on the manifold are $N_c\times N_c$ special unitary
matrices.  For any simply connected Lie group, the neighbourhood of any
point is isomorphic to the neighbourhood of identity. The neighbourhoods
are related by right-translating the elements in the neighbourhood of
identity by U \ie,
\beq
U^\prime=\exp\left(i\sum_{A=1}^N \omega_A T_A \right) U, 
\eeq{righttranslate}
where $\omega_A$ are real scalars. The matrix $U^\prime$ lies in the
neighbourhood of $U$ and $\exp\left(i\sum_A^N \omega_A T_A \right)$ lies
in the neighbourhood of identity for small values of $\omega_A$. The $T_A$
are $N_c\times N_c$ traceless hermitian matrices satisfying $\tr T_A T_B =
\delta_{A,B}$ and they are called the generators of $\sun$. Throughout
this paper, the letter $T$ (with or without super- and sub-scripts)
denotes a traceless hermitian matrix.  The group can be classified into
families of 1-parameter abelian subgroups characterized by different $T$,
each containing the elements $\exp\left(i r T\right)$, for real $r$. This
offers another way of finding the neighbourhood of $U$:
\beq
U^\prime=\exp\left(i r T \right) U.
\eeq{righttranslate2}

The $N$-tuple, $(\omega_1,\omega_2,\ldots,\omega_N)$, serves as a
coordinate chart for the neighbourhood of $U$, enabling us to find
derivatives of $f$ (or rather of $f\circ \phi^{-1}$).  Tangent vector space
at $U$, denoted by $\mathbf{T}(U)$, is the vector space of directional
derivatives tangent to the curves in $\mathbb{R}^N$ passing through
$U$. Usually, a tangent vector is defined as a directional derivative operator,
$\sum_{A=1}^N v_A \partial/\partial\omega_A$, with $v_A$ being real
scalars.  However, using the method of translation along a one parameter
subgroup, the tangent vectors in $\mathbf{T}(U)$, denoted by $\left[\delta
f/\delta U\right]_T$, become the ordinary total derivatives
\beq
\deriv{f}{T}{U}=\frac{d }{dr}f\left(e^{irT}U\right)\bigg|_{r=0} \equiv \tr\left(T \frac{\delta f}{\delta U}\right),
\eeq{tangentvector}
where we have also implicitly defined the gradient, $\delta f/\delta
U$.  In this method, $U$ varies along a curve, parametrized by $r$,
on the manifold and we find the derivative along the tangent to this
curve (one can think of this curve as being traced in the course of
a molecular dynamics trajectory).  This way of thinking is useful
for the case of $V(U)$, which is an $\sun$ valued function of $U$.
In this case, the curve traced by $U$ maps to another curve traced by
$V$ on its manifold. Let the tangents to the two curves at $U$ and $V$
be along the directions determined by $T$ and $T^\prime$ respectively.
Let $f(V)$ be an explicit function of $V$. The derivative of $f(V)$
with respect to variation in $U$ is a tangent vector at $V$ along the
direction $T^\prime$:
\beq
\deriv{f\left[V(U)\right]}{T}{U}= \frac{d }{dr}f\left[V(e^{irT}U)\right]\bigg|_{r=0}= \frac{d }{dr}f\left[e^{irT^\prime}V\right]\bigg|_{r=0}.
\eeq{chainrule}
\bef
\begin{center}
\includegraphics[scale=0.5]{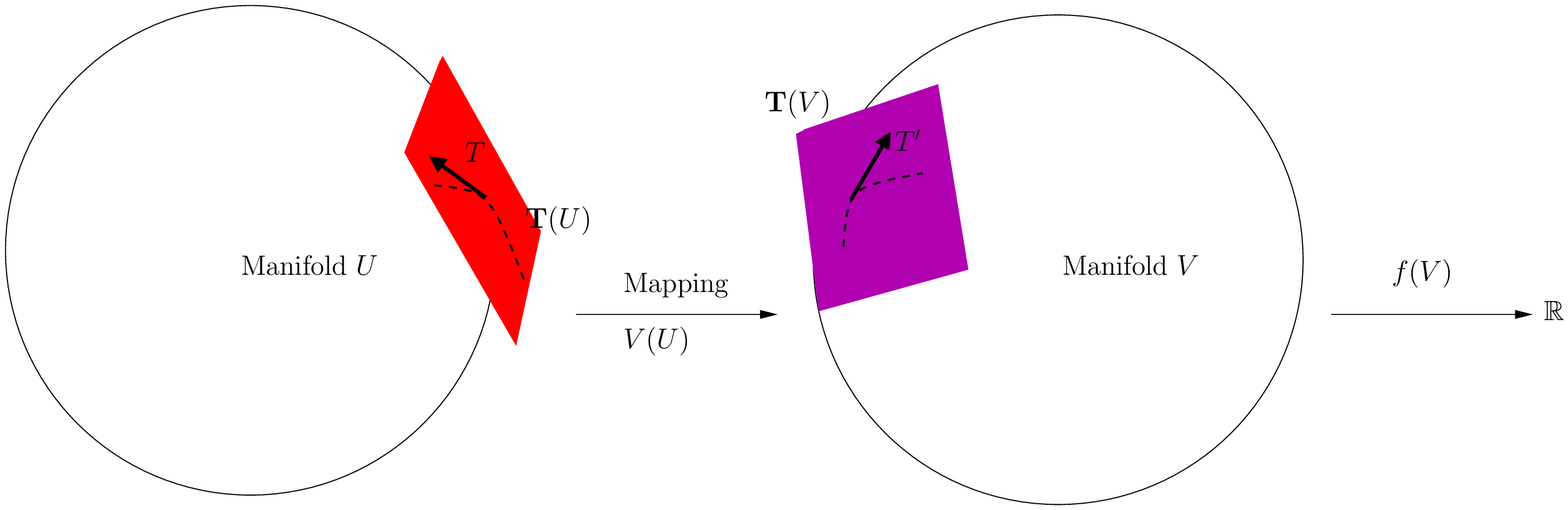}
\end{center}
\caption{Finding derivatives. The spheres on the left and the right
schematically represent the $\sun$ manifolds whose points represent $\sun$
matrices $U$ and $V$ respectively. $V$ is a function of $U$ \ie, there is
a mapping between the two manifolds. The red plane on the left represents
the tangent vector space, $\mathbf{T}(U)$, while the magenta plane on the
right is $\mathbf{T}(V)$. The matrix $U$ changes along the dashed curve
on the left and the tangent to the curve at $U$, in $\mathbf{T}(U)$,
is along the direction $T$ denoted by an arrow.  This variation in $U$
causes $V$ to vary along the dashed curve on the right.  The tangent to
the curve at $V$ is along the direction $T^\prime$, in $\mathbf{T}(V)$.
$f(V)$ is a real valued explicit function of $V$. As explained in the text, 
its derivative with respect to $U$ is now an ordinary total derivative in the
direction $T^\prime$. What remains is to find $T^\prime$.
}
\eef{tang}
The final expression in the above equation is the tangent vector
$\left[\delta f/\delta V\right]_{T^\prime}$.  This procedure is explained
schematically in \fgn{tang}.  To implement the chain rule,  now one has
to find an ordinary derivative in $\mathbf{T}(V)$. The problem is to
find the direction $T^\prime$, which we will show to be very simple.
This is to be contrasted with the usual method employed \cite{bmw}:
\beq
\deriv{f(V)}{T_A}{U}=\del{ }{u_A} f\left[V(e^{i u_A T_A}U)\right]\bigg|_{u=0}=\sum_{B}\del{v_B}{u_A}\del{ }{v_B}f\left[e^{i\sum_C v_C T_C}V\right]\bigg|_{u,v=0}.
\eeq{usual}
This requires writing
$\partial{f}/\partial{v_B}=\sum_{a,b}\left(\partial{f}/\partial{V^{ab}}\right)\left(\partial{V^{ab}}/\partial{v_B}\right)$,
which involves derivatives with respect to each matrix element of $V$.
In order to simplify further, we need derivatives of $U$ and $V$ with
respect to $r$. These derivatives are defined as
\beq
\dbd{U}{r}\equiv\dbd{e^{i r T} U}{r}\bigg|_{r=0}= i T U \quad\text{\rm and}\quad \dbd{V\left(U\right)}{r}=i T^\prime V.
\eeq{delu}
We shall write the above equation in short as $dU = i T U dr$ and call
$dU$ as the ``variation in $U$". The derivative $dU/dr$ is co-variant,
as the assignment of a matrix $U$ to each point on the manifold is
unique up to a global transformation $G^\dagger U G$ with $G\in \sun$,
thereby not requiring affine connections which are necessary when local
transformations exist.

We end this section by noting a simple identity that will be very useful
in \scn{force}: for any two matrices $M$ and $N$, projection to the
space spanned by the generators satisfies
\beq
\re\tr(M\pa{N})=\re\tr(\pa{M}N).
\eeq{projid}

\section{Calculation of force}\label{sec:force}
A trajectory of HMC \cite{hmc} consists of an initial refreshment
of momenta, $\Pi_{x,\mu}$, followed by classical evolution of gauge
fields and their conjugate momenta through the simulation time, $\tau$,
according to a fictitious Hamiltonian, $H$. At the end of a trajectory,
an acceptance step corrects for the discretisation error in the symplectic
integrator. By definition, $\Pi_{x,\mu}$ determines the direction in
which $U_{x,\mu}$ should evolve (refer \eqn{delu}). The Hamiltonian for
staggered fermions is \cite{ralgo}
\beq
H=\frac{1}{2}\sum_{x,\mu} \tr\Pi_{x,\mu}^2-S_g(U)-S_f\quad{\rm and}\quad S_f=\frac{N_f}{8}{\rm Tr}\ln \left(D^\dagger D\right),
\eeq{hamil}
where  $S_g(U)$ is the gauge action and $S_f$ is the fermion action. The
momenta are evolved such that $H$ is conserved. These conditions give
the equations of motion
\beq
\frac{d U_{l,\mu}}{d\tau}=i\Pi_{l,\mu}U_{l,\mu}\quad{\rm and}\quad \frac{d \Pi_{l,\mu}}{d\tau}=\frac{\delta S_f}{\delta U_{l,\mu}}+\frac{\delta S_g}{\delta U_{l,\mu}}.
\eeq{eqp}
$\delta S_f/\delta U$ and $\delta S_g/\delta U$ are called the fermion
and gauge forces respectively.  The fermion force in the case of the
standard staggered action was derived in \cite{ralgo} and we give a
slightly different derivation of it in \scn{thinforce}. In \scn{hexforce},
we derive the fermion force for HEX improved staggered action.

\subsection{Fermion force for standard staggered action}\label{sec:thinforce}
The Dirac operator used in the standard staggered fermion action 
is a function of thin-links, $U_{x,\mu}$ (refer \eqn{stagdef}). 
When the link at site $l$ varies in its tangent vector space by
$dU_{l,\mu}=i dr T U_{l,\mu}$, the fermion force is given by 
\beq
\frac{8}{N_f}\deriv{S_f}{T}{U_{l,\mu}}=\tr\left(D^{-1}\delt{D}{T}{r}+ \text{\rm h.c.}\right).
\eeq{fermforce}
The derivative with respect to $U_{l,\mu}$ has been converted to an
ordinary derivative with respect to $r$ in its tangent vector space using
\eqn{delu}.  By using the definition of $D$ given in \eqn{stagdef},
the ordinary derivative becomes
\beq
\delt{D_{x,y}}{T}{r}=\frac{i}{2}\left(T U_{x,\mu}\delta_{x,y-\mu}\delta_{x,l}+ U^\dagger_{x-\mu,\mu}T\delta_{x,y+\mu}\delta_{x-\mu,l}\right),
\eeq{cdirac}
where we have used \eqn{delu}. Having found the gauge derivative of $D$
in one simple calculation, \eqn{fermforce} simplifies to
\beq
\frac{16}{N_f}\deriv{S_f}{T}{U_{l,\mu}}=\tr\left[iT\left(U_{l,\mu}D^{-1}_{l+\mu,l}+D^{-1}_{l,l+\mu}U^\dagger_{l,\mu}\right)+\rm{h.c.}\right]\equiv \re\tr\left[iT \Sigma_{l,\mu}\right],
\eeq{thinforce}
where we have factored out $T$ and collected the remaining terms as
$\Sigma_{l,\mu}$. Taking the real part is superfluous. However we do so
anticipating the simplifications in the next subsection.  The fermion
force follows by finding these derivatives along the generators.  $D^{-1}$
is usually evaluated by inserting a stochastic estimator of identity
\cite{ralgo}, but that does not concern our calculation.

\subsection{Fermion force for HEX smeared staggered action}\label{sec:hexforce}
To avoid unnecessary complications, we restrict ourselves to one
level of HEX improvement. The fermion action, $S_f$, is now an explicit
function of the HEX smeared links, $V^{(3)}$ (refer \eqn{hexschematic}).
Having demonstrated the method of finding the ordinary derivative in
tangent vector space for the case of the standard staggered action,
we now demonstrate how it greatly simplifies the implementation of the
chain rule required for hypercubic smearing schemes.

\subsubsection{Essential simplification}
The HEX smeared links in the $n$-th sub-level, $V^{(n)}_x$, are explicit
functions of both thin-links and the smeared links in the $(n-1)$-th
sub-level. Any variation, $dU_{l,\mu}=i dr T_A U_{l,\mu}$, in the
thin-link at site $l$ causes variations $dV^{(n)}_x= i dr T^{(n)}_x
V^{(n)}_x$ in the smeared links (that are within the hypercube containing
$U_{l,\mu}$). It is to be noted that we have dropped directional
indices for the sake of brevity and ease of generalization to various
sub-levels. Applying the product rule to \eqn{hexschematic} and rewriting
it in the form $i dr T^{(n)}_x V^{(n)}_x$, leads to the expression
\beq
T^{(n)}_x= V^{(n)}_x U^\dagger_x T_A U_x V^{(n)\dagger}_x\delta_{x,l}-i\dbd{\exp(A^{(n)}_x)}{r} U_x V^{(n)\dagger}_x,
\eeq{tmdir}
where we have replaced $\exp(A)$ by $VU^\dagger$ to get to the
above expression.  Thus, we have determined the directions in
$\mathbf{T}(V^{(n)})$, along which ordinary derivatives are to be found.

\subsubsection{Recurrence Relation}
The rest of the calculation proceeds backwards from the third sub-level
by merely finding the ordinary derivatives with respect to $r$.  Using the
chain-rule (refer \eqn{chainrule}),
\beq
\deriv{S_f}{T_A}{U_{l,\mu}}=\sum_{x,\nu}\deriv{S_f}{T^{(3)}_{x,\nu}}{V^{(3)}_{x,\nu}}.
\eeq{leibnitz}
Since the HEX improved Dirac operator is obtained from the standard
staggered Dirac operator by replacing $U$ with $V^{(3)}$, we can simplify
$\delta S_f/\delta V^{(3)}$ by borrowing results from \scn{thinforce}.
After such replacements in \eqn{thinforce},
\beq
\frac{16}{N_f}\deriv{S_f}{T_A}{U_{l,\mu}}=\sum_{x,\nu}\re\tr\left[i T^{(3)}_{x,\nu}\Sigma^{(3)}_{x,\nu}\right],
\eeq{st3strt}
where $\Sigma^{(3)}_{x,\nu}$ is the HEX version of $\Sigma$.
By using \eqn{tmdir} for $T^{(3)}$ and defining $\Sigmab^{(n)}\equiv
V^{(n)\dagger}\Sigma^{(n)}$,
\beq
\frac{16}{N_f}\deriv{S_f}{T_A}{U_{l,\mu}}=\re\tr\left[i T_A U_{l,\mu} \Sigmab^{(3)}_{l,\mu} V^{(3)}_{l,\mu} U^\dagger_{l,\mu}\right] +\sum_{x,\nu}\re\tr\left[U_{x,\nu}\Sigmab^{(3)}_{x,\nu}\dbd{\exp(A^{(3)}_{x,\nu})}{r}\right].
\eeq{st3ep2}
The next step is to reduce the derivative of $\exp(A^{(3)})$ to
a derivative of its exponent. Making a power series expansion of
$\exp(A^{(3)})$, one would expect that a matrix $\Lambda^{(3)}$ can be
defined such that,
\beq
\re\tr\left[\Lambda^{(3)}_{x,\nu} \dbd{A^{(3)}_{x,\nu}}{r}\right]\equiv\re\tr\left[U_{x,\nu}\Sigmab^{(3)}_{x,\nu}\dbd{\exp(A^{(3)}_{x,\nu})}{r}\right].
\eeq{st3ep3}
For the case of ${\rm SU}(3)$, the Cayley-Hamilton theorem leads to such
a simplification by the expansion of $d\left[\exp(A)\right]$ in terms
of $A, A^2$ and $dA$. This was done in \cite{ST} and using their result,
\beq
\Lambda_x=\tr(U_{x}\Sigmab_{x} B_1)A_{x}+\tr(U_{x}\Sigmab_{x} B_2)A_{x}^2+
f_1 U_{x}\Sigmab_{x} + f_2 A_{x} U_{x}\Sigmab_{x} + f_2 U_{x}\Sigmab_{x} A_{x},
\eeq{deflambda1}
where $B_i=\sum_{j=0}^2 b_{ij}A^j$. The coefficients $f_i$ and $b_{ij}$
are complex scalar functions of the eigenvalues of $A$ \cite{ST}.
Using \eqn{projid} and the definition of $A$ in \eqn{hexschematic}, it
is possible to write $\re\tr\left(\Lambda dA\right)=\re\tr\left(\Lambda
d\pa{CU^\dagger}\right)$ as
\beq
\re\tr\left[\Lambda^{(n)}_x\frac{d}{dr}\pa{C^{(n)}_x U^\dagger_x}\right]=\re\tr\left[\pa{\Lambda^{(n)}_x}\dbd{C^{(n)}_x}{r} U^\dagger_x\right] -\re\tr\left[i T_A\pa{\Lambda^{(n)}_x}C^{(n)}_x U^\dagger_x\right]\delta_{x,l}.
\eeq{st3ep4}
Using \eqn{hxdef}, we can schematically write $dC=d(V)VV+Vd(V)V+VVd(V)$
with the smeared links, $V$, which are one sub-level below.  Each $dV$
gives rise to a direction vector $T$, which can be cyclically permuted
as the first term due to presence of the trace.  Since the spatial
and directional indices are summed over in \eqn{st3ep2}, these are
are dummy indices and we can factor out a $T$ for each site $x$ and direction.
Reserving this calculation for \apx{sigma}, we define a matrix $\Sigma$ as
\beq
\sum_x\re\tr\left[i T^{(n-1)}_x \Sigma^{(n-1)}_x\right]\equiv\sum_x\re\tr\left[Z^{(n)}_x \dbd{C^{(n)}_x}{r}\right]\quad{\rm where}\quad Z^{(n)}_x=U_x^\dagger\pa{\Lambda^{(n)}_x}.
\eeq{sigmadef}
The matrix $\Sigma$ arises naturally in this method leading to a very
simple calculation (given in \apx{sigma}) unlike the conventional
method in \cite{bmw} where derivatives of staple with respect to matrix
elements of $V$ are required. Using \eqn{st3ep2}, \eqref{st3ep4} and
\eqref{sigmadef}, after putting back the indices for the third sub-level,
\beq
\begin{split}
\frac{16}{N_f}\deriv{S_f}{T_A}{U_{l,\mu}}= {}&\re\tr\left[i T_A U_{l,\mu}\left\{\Sigmab^{(3)}_{l,\mu} V^{(3)}_{l,\mu} - Z^{(3)}_{l,\mu} C^{(3)}_{l,\mu}\right\}U^\dagger_{l,\mu}\right]+\sum_{x,\mu,\nu}{}^\prime\re\tr\left[i T^{(2)}_{x,\mu;\nu}\Sigma^{(2)}_{x,\mu;\nu}\right].
\end{split}
\eeq{recur}
$\sum{}^\prime$ indicates that none of the directions are equal.
In the above equation, the second term in the right hand side has the
same form as \eqn{st3strt}, thereby giving us a recurrence relation.
Along with the conditions that $V^{(0)}=U$ and $C^{(0)}=0$, the rest of
the terms can be written down. The equation for the fermion force becomes
\beq
\begin{aligned}
\frac{16}{i N_f}\frac{\delta{S_f}}{\delta U_{l,\mu}}={}&\pa{U_{l,\mu}\left\{\Sigmab^{(3)}_{l,\mu} V^{(3)}_{l,\mu} - Z^{(3)}_{l,\mu} C^{(3)}_{l,\mu}\right\}U^\dagger_{l,\mu}} +\\
&\sum_{\nu}{}^\prime\pa{U_{l,\mu}\left\{\Sigmab^{(2)}_{l,\mu;\nu} V^{(2)}_{l,\mu;\nu} - Z^{(2)}_{l,\mu;\nu} C^{(2)}_{l,\mu;\nu}\right\}U^\dagger_{l,\mu}} +\\
&\sum_{\nu,\rho}{}^\prime\pa{U_{l,\mu}\left\{\Sigmab^{(1)}_{l,\mu;\nu\rho} V^{(1)}_{l,\mu;\nu\rho} - Z^{(1)}_{l,\mu;\nu\rho} C^{(1)}_{l,\mu;\nu\rho}\right\}U^\dagger_{l,\mu}} +\\
&\sum_{\nu,\rho,\eta}{}^\prime\pa{U_{l,\mu}\Sigmab^{(0)}_{l,\mu;\nu\rho\eta}} .
\end{aligned}
\eeq{master}
If the action is improved by multiple levels of HEX, then $C^{(0)}$
is the staple constructed out of the smeared links one level below,
which is $V^{(0)}$. Thus, the recurrence relation is easily extended to
multiple levels.

\section{Memory optimized implementation of hypercubic schemes}\label{sec:memopt}
\bef
\begin{center}
\includegraphics[scale=0.20]{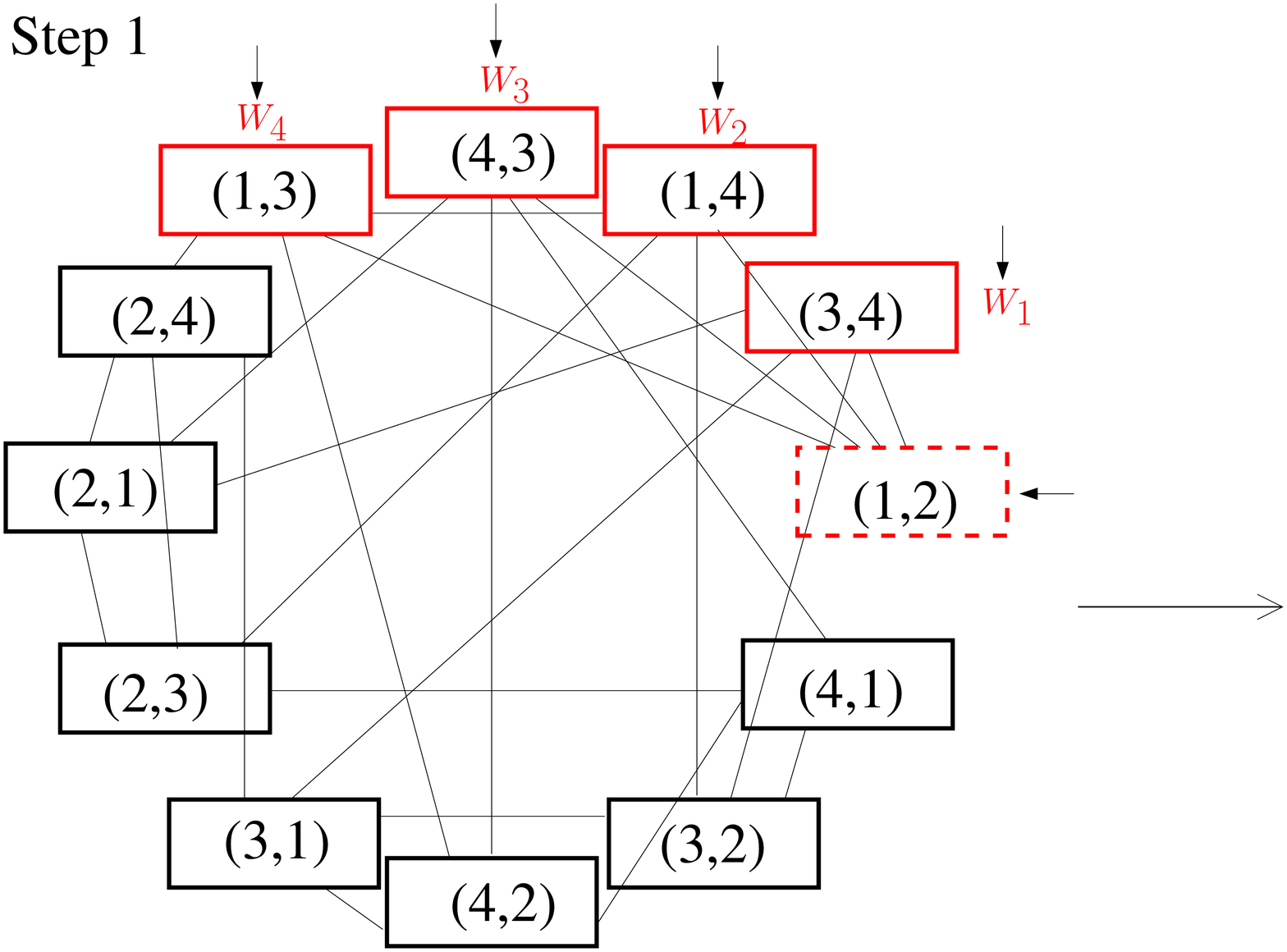}
\includegraphics[scale=0.20]{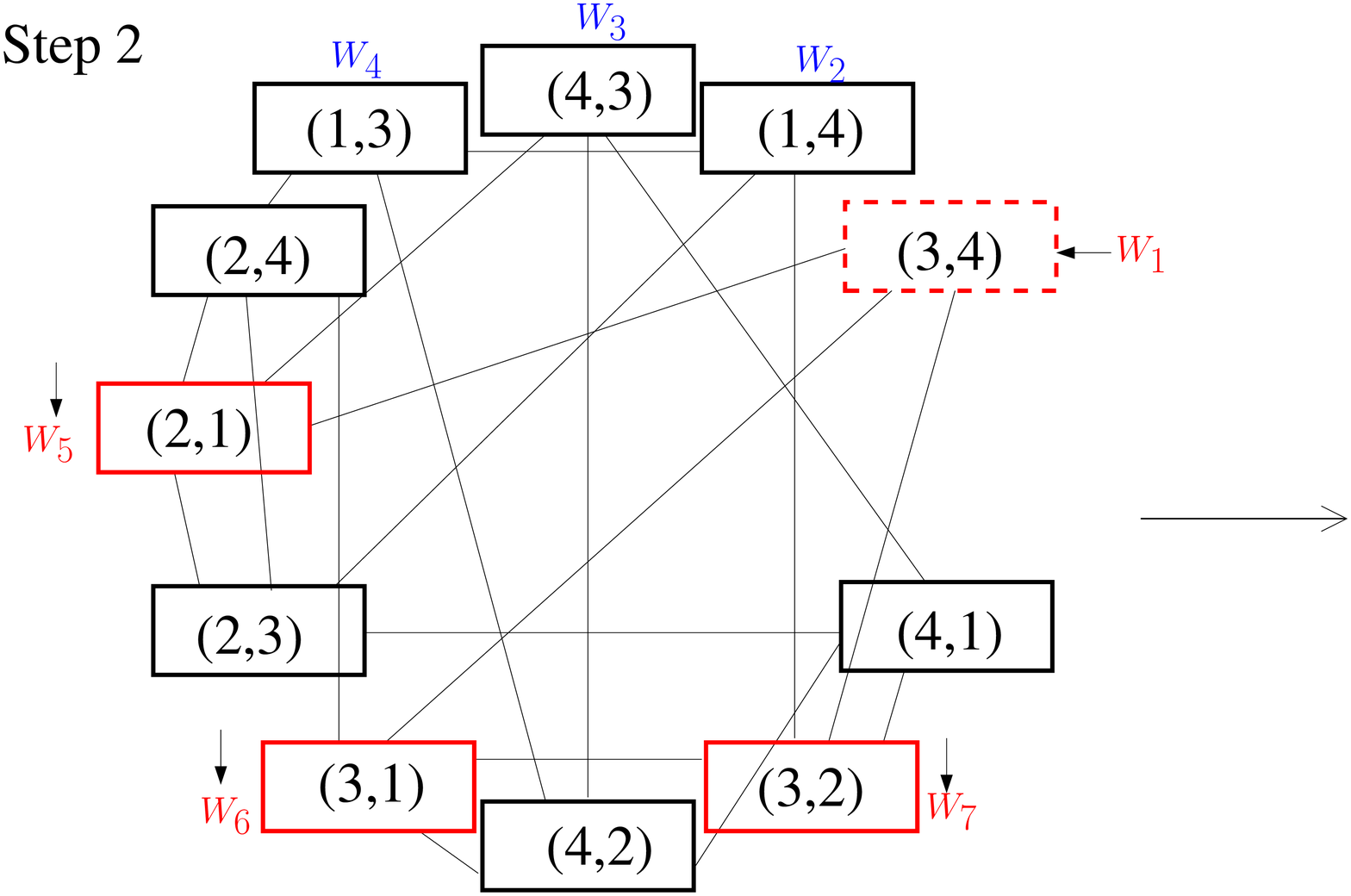}
\includegraphics[scale=0.20]{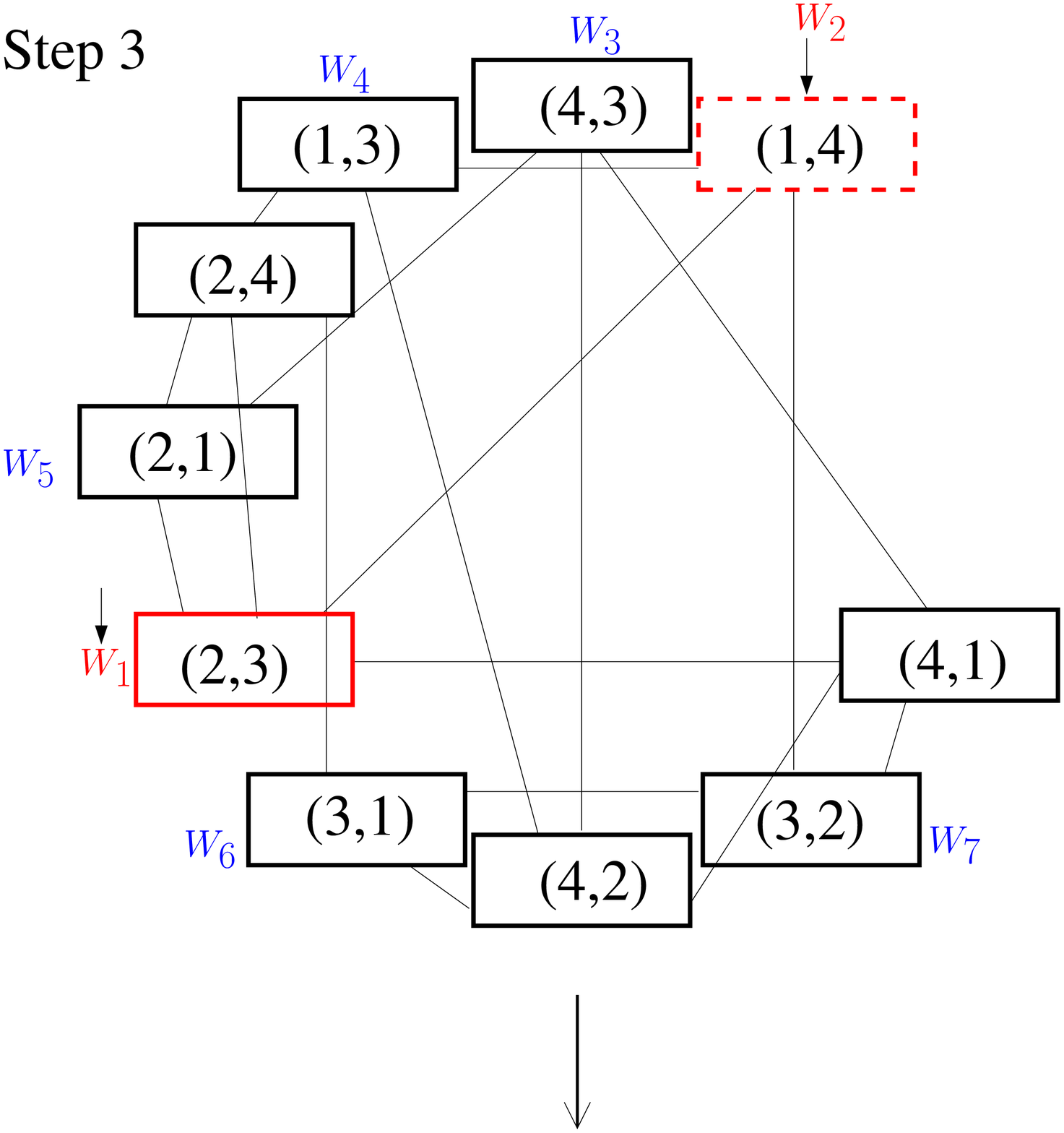}

\includegraphics[scale=0.20]{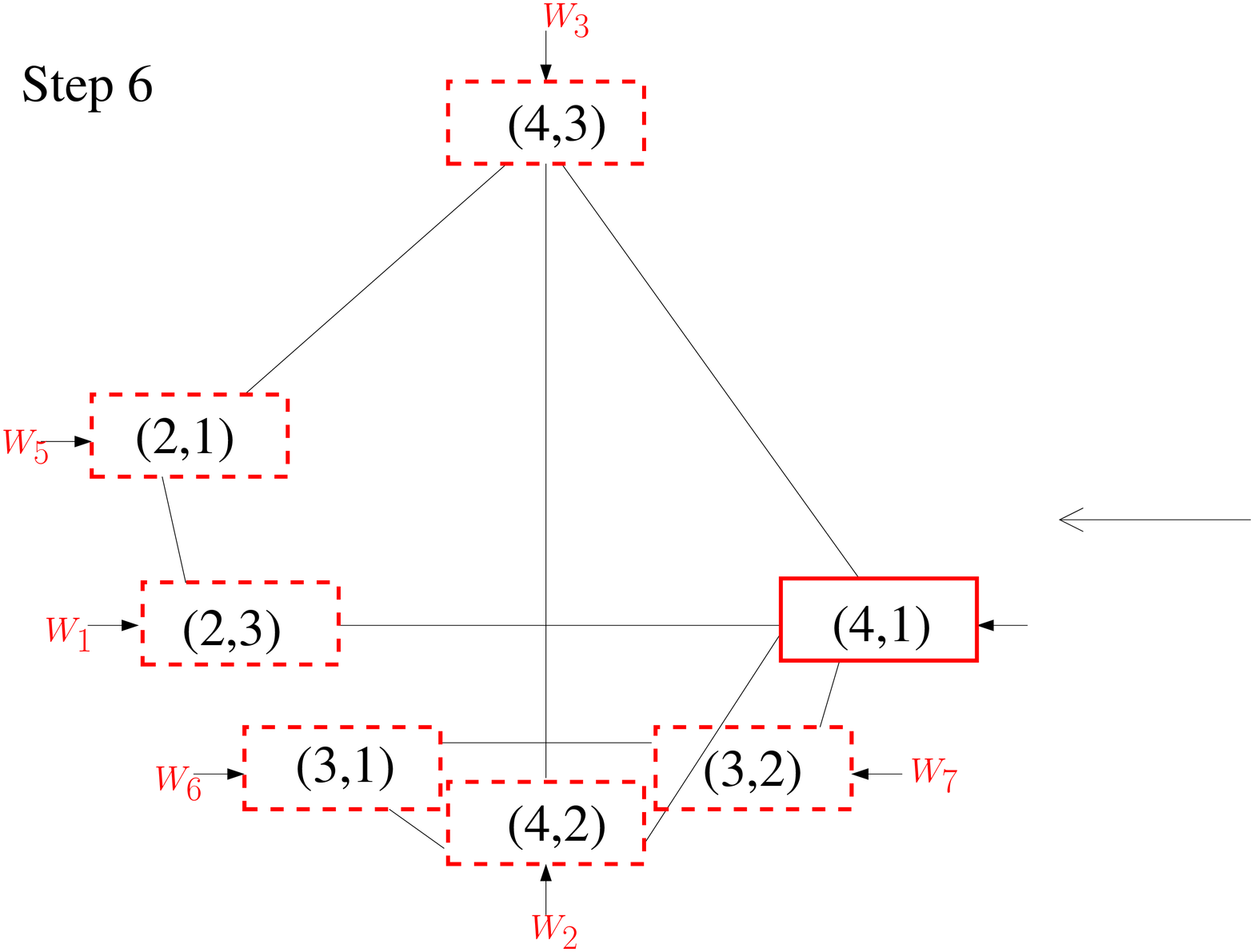}
\includegraphics[scale=0.20]{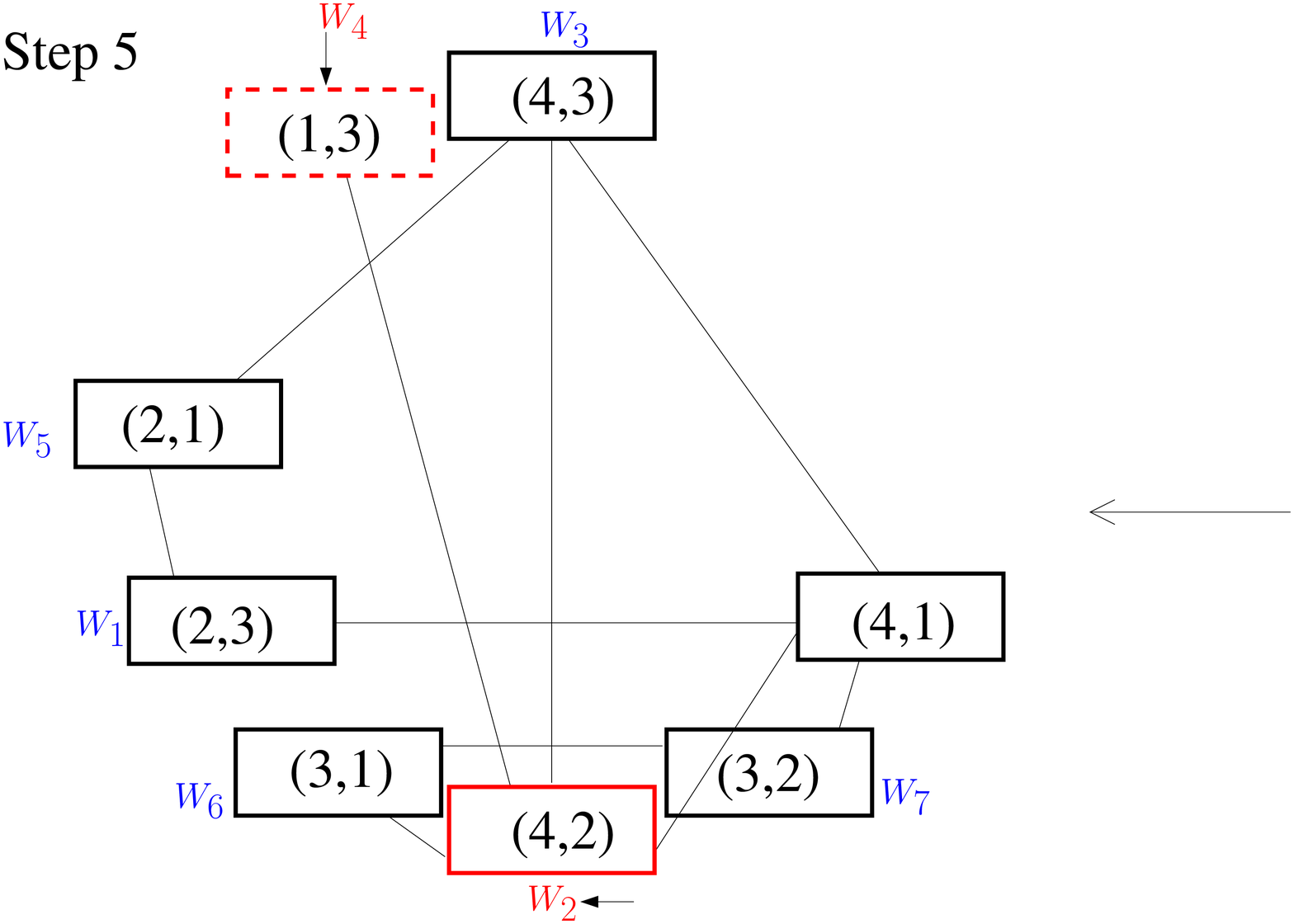}
\includegraphics[scale=0.20]{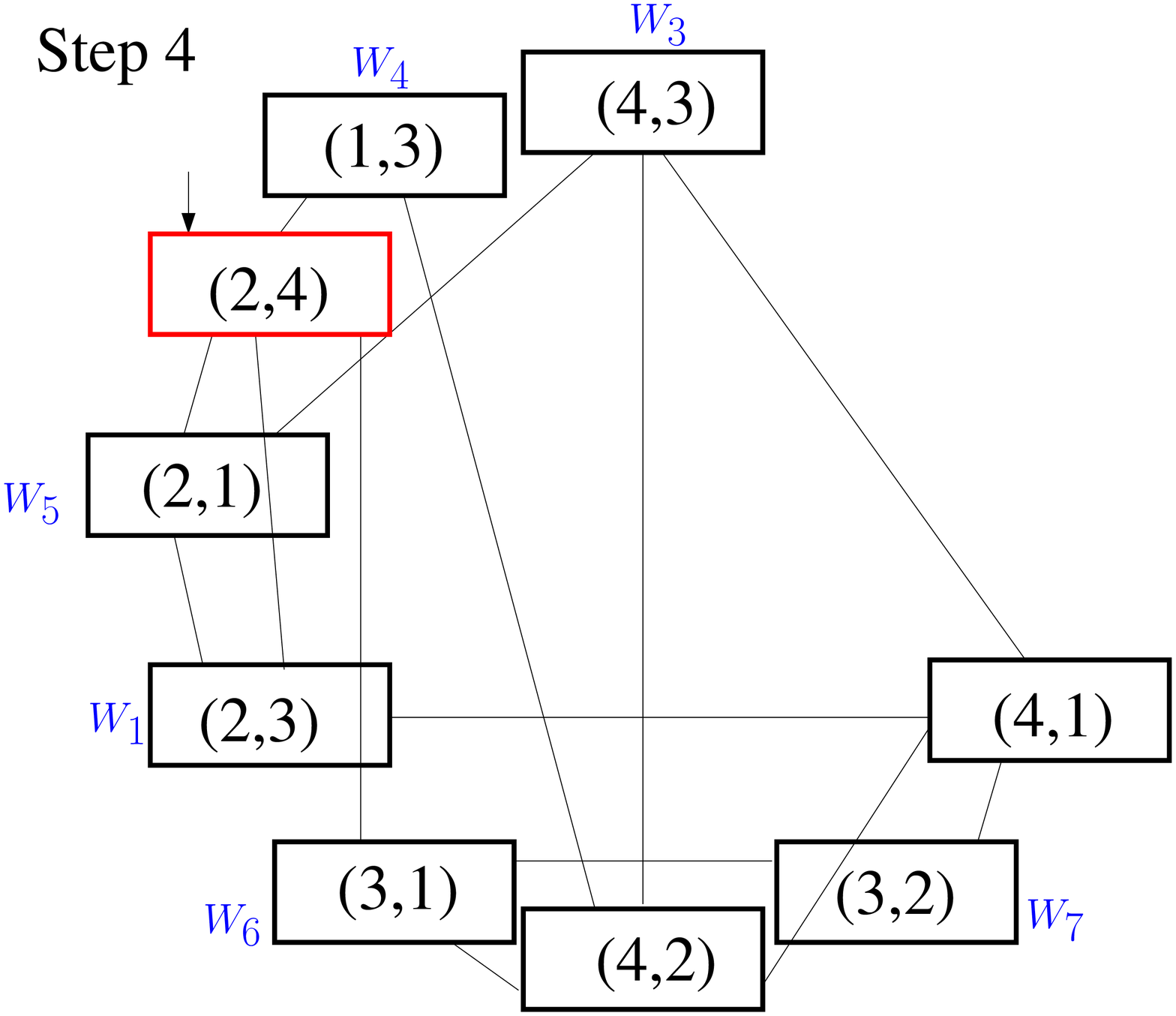}
\end{center}
\caption{Graphical illustration of memory optimized algorithm for
the second sub-level of hypercubic schemes.  The vertices $(\mu,\nu)$
are the $9 V_{lat}$ sized arrays, $V_{\mu;\nu}$, which stores $3\times
3$ matrices at each lattice point.  Two vertices, $(\mu,\nu)$
and $(\rho,\eta)$, are connected by an edge (\ie, adjacent) if the
construction of $V^{(2)}_{\mu;\nu}$ depends on $V^{(1)}_{\rho;\eta}$
(see \eqn{hxdef}), in which case $V^{(2)}_{\rho;\eta}$ also depends
on $V^{(1)}_{\mu;\nu}$.  There are 7 working arrays, $W_i$, which are
indicated near some vertices and each of them are of size $9V_{lat}$.
The algorithm proceeds from step 1 to 6 as shown in the figure. At
the beginning of step 1, $(\mu,\nu)$ contains $V^{(1)}_{\mu;\nu}$
at all lattice points.  In each step, the red colored arrays are the
ones which are relevant and the blue ones are dormant.  Each step consists
of the following operations.  A short arrow with nothing at its tail,
$\rightarrow W_i$ or $\rightarrow$\fbox{$(\mu,\nu)$}, indicates that
$V^{(2)}_{\mu;\nu}$ is computed using $V^{(1)}_{\rho;\eta}$ contained in
the vertices adjacent to $(\mu,\nu)$ and stored in the array it points
to. $W_i\rightarrow$\fbox{$(\mu,\nu)$} means that data is copied from
$W_i$ to $V_{\mu;\nu}$. Solid or dashed red vertices indicate the order of
these operations: the operations at the solid red vertices are done before
the dashed ones, and amongst the solid vertices there is no hierarchy.
Once $V_{\mu;\nu}$ is updated with $V^{(2)}_{\mu;\nu}$, the vertex and
the edges attached to it are removed from the successive steps.}
\eef{opthyp}

A code implementation of \eqn{hxdef} requires arrays $U_{\mu}$ and
$V_{\mu;\nu}$ to store thin-links and the subsequent smeared links in
various sub-levels respectively. Each of these arrays store $3\times
3$ matrices at all lattice points and hence each of them is of size $9
V_{lat}$. Since each step consists of updates done at all lattice points,
we do not show the position indices of these arrays.  In addition to them,
we require arrays $W_i$ for $1\le i \le N_w$. These are required in order
to provide work space to enable updating the same array $V_{\mu;\nu}$
with smeared links, when going from one sub-level to the next. Each of
these arrays are again of size $9 V_{lat}$. The problem is to minimize
$N_w$. The first sub-level is easy to implement as it requires only
thin-links, and the $V_{\mu;\nu}$ are empty to begin with. This step does
not require any working arrays.

The smeared links, $V^{(1)}_{\sigma;\mu\nu}$, used in the second
sub-level can also be written as $V^{(1)}_{\sigma;\eta}$, where $\eta$
is the direction orthogonal to $\sigma$, $\mu$ and $\nu$. With this
observation, it is clear that both $V^{(1)}_{\sigma;\mu\nu}$ and
$V^{(2)}_{\sigma;\eta}$ can be stored in the same array of the form
$V_{\sigma;\eta}$.  The brute force implementation of this sub-level
would require $N_w=12$, as two copies of $V$ are required: one to store
$V^{(1)}$ and the other for $V^{(2)}$.

A graphical representation of an algorithm to compute the second
sub-level, so that $N_w$ is reduced to 7, is given in \fgn{opthyp}.
The vertices $(\mu,\nu)$ stand for the arrays $V_{\mu;\nu}$.  Two
vertices $(\mu,\nu)$ and $(\rho,\eta)$ are adjacent, if \eqn{hxdef} for
$V^{(2)}_{\mu;\nu}$ involves $V^{(1)}_{\rho;\eta}$. If this is true, then
by observation, $V^{(2)}_{\rho;\eta}$ also involves $V^{(1)}_{\mu;\nu}$
and hence this graph is not directed.  At the beginning of the second
sub-level, $(\mu,\nu)$ contains $V^{(1)}_{\mu;\nu}$ at all lattice
points. $(\mu,\nu)$ cannot be updated with $V^{(2)}_{\mu,\nu}$ until the
$V^{(2)}_{\rho;\eta}$ corresponding to all the adjacent vertices have
been computed.  The way to proceed becomes clear through the circular
embedding of the graph.  The algorithm is detailed in the caption of
\fgn{opthyp}. Here, we give a walk-through of the first two steps as
follows.  In the first step, we arbitrarily pick a vertex --- we choose
$(1,2)$ in this example. The calculation of $V^{(2)}_{3;4}, V^{(2)}_{1;4},
V^{(2)}_{4;3}$ and $V^{(2)}_{1;3}$ requires $V^{(1)}_{1;2}$. Therefore,
we first find $V^{(2)}$ for these adjacent vertices and store them in
the working arrays $W_1, W_2, W_3$ and $W_4$.  Now, we are free to update
$(1,2)$ with $V^{(2)}_{1;2}$. This requires $V^{(1)}_{3;4}, V^{(1)}_{1;4},
V^{(1)}_{4;3}$ and $V^{(1)}_{1;3}$, which still remain untouched in the
adjacent vertices. This is the reason for the specific order of these
updates described in the figure. Once $(1,2)$ is updated, no other
vertex requires it and it gets disconnected from the graph. The second
step proceeds similarly with respect to the vertex $(3,4)$. However,
at the end of the second step, $(3,4)$ is updated to $V^{(2)}_{3,4}$,
which was stored in $W_1$ during the first step.  The rest of the steps
of this algorithm proceed by repeating this procedure for a specific
sequence of vertices, as shown in \fgn{opthyp}, such that only 7 working
arrays are required at any point of the algorithm. The total memory cost
for the working space in terms of array size is $63 V_{lat}$ compared
to $108 V_{lat}$ in the brute-force method.  This cost can be further
reduced by appealing to unitarity: only two rows of $W_i$ at each lattice
point are required. This reduces the memory cost further to $42 V_{lat}$.

The third sub-level is again straight-forward. Only 4 working arrays
are needed to compute $V^{(3)}_{\mu}$ (due to the 4 values of $\mu$).

\section{Summary}
In this paper, we presented a simple method for implementing the chain
rule to calculate fermion forces for the Hybrid Monte Carlo algorithm
with HEX improved staggered action. It is based on the fact that the
derivative of the action with respect to a thin-link can be written as
the ordinary derivatives with respect to a real parameter in each of
the tangent vector spaces of the smeared links. This way, we were able
to avoid finding the derivatives with respect to all matrix elements of
thin-links, as done in the literature \cite{kamleh,bmw}. This has the
obvious advantage of making the calculation very simple (\scn{hexforce}).
In \scn{memopt}, we gave an implementation of hypercubic schemes with
less than half the memory requirement for work space when compared to
the brute force implementation.

I would like to thank Prof. Sourendu Gupta for the discussions and for
the idea of using the methods of differential geometry. I also 
thank him and Rahul Dandekar for careful reading of the manuscript.

\appendix

%%%%%%%%%%%%%%%%%%%%%%%CHECKPOINT-#calsig
\section{Calculation of $\Sigma$}\label{sec:sigma}
The definition of $\Sigma^{(3)}_{x,\mu}$ is given in \eqn{st3strt}
and it involves the thin-link force with the replacement of thin-links
with the smeared links, $V^{(3)}$. For $n=0, 1$ and $2$, $\Sigma^{(n)}$
is defined by the relation
\beq
\sum_x \re\tr\left[i T^{(n)}_x \Sigma^{(n)}_x\right]\equiv\sum_x\re\tr\left[Z^{(n+1)}_x\dbd{C^{(n+1)}_x}{r}\right].
\eeq{sigdef2}
From \eqn{hxdef}, $C^{(n+1)}$ is constructed out of products of
three $V^{(n)}$.  Using the product rule and then the cyclicity of the
trace, we take all $dV$ to the front and $dV^\dagger$ to the end. Due to
summation over all spatial indices, we translate each term such that $dV$
or $dV^\dagger$ in all the terms are at position $x$. Due to summation
over all the directional indices, these are dummy indices and can be
interchanged. $dV$ and $dV^\dagger$ bring a factor of $T^{(n)}_x$ to the
front by this construction.  We give the result for the second sub-level:
\beq
\begin{aligned}
\frac{6}{\epsilon_3}\Sigma^{(2)}_{x,\nu;\sigma}= {}&V^{(2)}_{x,\sigma;\nu}V^{(2)}_{x+\sigma,\nu;\sigma} V^{(2)\dagger}_{x+\nu,\sigma;\nu}Z^{(3)}_{x,\nu}
+V^{(2)}_{x,\sigma;\nu}V^{(2)\dagger}_{x-\nu+\sigma,\nu;\sigma}Z^{(3)}_{x-\nu,\sigma}V^{(2)}_{x-\nu,\nu;\sigma}\\
&-Z^{(3)}_{x-\nu,\nu}V^{(2)}_{x-\nu,\sigma;\nu}V^{(2)}_{x-\nu+\sigma,\nu;\sigma}V^{(2)\dagger}_{x,\sigma;\nu}
-V^{(2)}_{x,\nu;\sigma}V^{(2)}_{x+\nu,\sigma;\nu}Z^{(3)}_{x+\sigma,\nu}V^{(2)\dagger}_{x,\sigma;\nu}\\
&+V^{(2)}_{x,\sigma;\nu}V^{(2)}_{x+\sigma,\nu;\sigma}Z^{(3)}_{x+\nu,\sigma}V^{(2)\dagger}_{x,\nu;\sigma}
+V^{(2)}_{x,\sigma;\nu}Z^{(3)}_{x+\sigma-\nu,\nu}V^{(2)\dagger}_{x-\nu,\sigma;\nu} V^{(2)}_{x-\nu,\nu;\sigma}.
\end{aligned}
\eeq{sig2}
It is straight forward to use appropriate directional
indices to get $\Sigma^{(1)}_{x,\nu;\sigma\rho}$ and
$\Sigma^{(0)}_{x,\nu;\sigma\rho\eta}$.  Using these, we can calculate
$\Sigmab=V^\dagger \Sigma$.

%%%%%%%%%%%%%%%%%%%%%%%%%%%%%%%%%%%%%%%%%%%%%%%%%%%%%%%%%%%%%%%%%%

\end{document}